\documentclass[prl,superscriptaddress,showpacs,twocolumn]{revtex4}
\usepackage{graphicx} 
\usepackage{amsmath}

\newcommand{\chalmersTF}{Department of Applied Physics,
Chalmers University of Technology,
SE-41296 G\"{o}teborg, Sweden}
\newcommand{\chalmersMC}{Department of Microtechnology and Nanoscience, MC2,
Chalmers University of Technology,
SE-41296 G\"{o}teborg, Sweden}
\newcommand{\camd}{Center for Atomic-scale Materials Design, Department of Physics, 
Technical University of Denmark,
DK-2800 Kongens Lyngby, Denmark}
\newcommand{\rutgers}{Department of Physics and Astronomy, Rutgers University,
Piscataway, NJ 08854-8019, USA}
\newcommand{\GUfysik}{Department of Physics, G\"oteborg University, SE-41296 G\"oteborg, Sweden}

\begin{document}
\title{Evaluation of New Density Functional with Account of van der Waals Forces 
by Use of Experimental H$_2$ Physisorption Data on Cu(111)}

\author{Kyuho Lee}\affiliation{\rutgers}
\author{Andr\'e K.~Kelkkanen}\affiliation{\camd}\affiliation{\chalmersMC}
\author{Kristian Berland}\affiliation{\chalmersMC}
\author{Stig Andersson}\affiliation{\GUfysik}
\author{David~C.~Langreth}\affiliation{\rutgers}
\author{Elsebeth Schr\"oder}\affiliation{\chalmersMC} 
\author{Bengt I.~Lundqvist}\affiliation{\camd}\affiliation{\chalmersMC}\affiliation{\chalmersTF}
\author{Per Hyldgaard}\affiliation{\chalmersMC}
\date{\today}


\begin{abstract}
Detailed experimental data for physisorption potential-energy 
curves of H$_2$ on low-indexed faces of Cu challenge theory.
Recently, density-functional theory has been developed to also account for nonlocal
correlation effects, including van der Waals forces.
We show that one functional, 
denoted vdW-DF2, gives a potential-energy curve promisingly close to 
the experiment-derived physisorption-energy curve. The comparison also 
gives indications for further improvements of the functionals.
\end{abstract}
\maketitle
Density-functional theory (DFT) gives in principle exact descriptions 
of stability and structure of electron systems, but in practice 
approximations have to be made to describe electron exchange and correlation 
(XC)~\cite{Kohn,KohnSham}. 
Evaluation of XC functionals is commonly done by comparison with results 
from other accurate electron-structure theories or by comparing
with relevant experiments, typically providing test numbers only 
for one or two measurables. This Letter illustrates the advantages of 
a third approach, which builds the experiment-theory calibration  
 based on extensive experimental data, in this case a 
full physisorption potential derived from surface-physics measurements.

In the physisorption regime, resonant elastic backscattering-diffraction 
experiments provide a detailed quantitative knowledge. Here, data 
obtained for H$_2$  and D$_2$  on Cu 
surfaces~\cite{andersson1993,perandersson1993,persson2008,Roy} 
are used as a demanding benchmark for the performance 
of adsorbate potential-energy curves (PECs) calculated with a 
nonempirical theory for extended systems. Density functionals that 
aspire to account for nonlocal electron-correlation effects, including 
van der Waals (vdW) forces, can then be assessed. In particular, we study 
the vdW-DF method \cite{dion2004,langreth2005,thonhauser2007,LeeEtAl10} 
and show that calculations with versions of it provide a promising 
description of the physisorption potential for H$_2$ on the Cu(111) 
surface and that the most recent one, vdW-DF2~\cite{LeeEtAl10}, is more 
accurate than the first one~\cite{dion2004,langreth2005,thonhauser2007} 
and other tested functionals.

Sparse matter is abundant. Dense matter, also abundant, is since long
successfully described by DFT. The recent extensions of DFT functionals 
to regions of low electron density, where the ubiquitous vdW forces 
are particularly relevant, render DFT useful also for sparse matter. 
In the vdW-DF functional the vdW interactions and correlations are 
expressed in the density $n(\mathbf{r})$ as a truly nonlocal 
six-dimensional integral~\cite{dion2004,langreth2005,thonhauser2007}. 
Its key ingredients are (i) its origin in the adiabatic connection 
formula~\cite{PerdLangrI,GunnLund,PerdLangrII}, 
(ii) an approximate coupling-constant integration, (iii) the use of an 
approximate dielectric function in a single-pole form, (iv) which is 
fully nonlocal and satisfies known limits, sum rules, and invariances, 
and (v) whose pole strength is determined by a sum rule and whose pole 
position is scaled to give the approximate gradient-corrected electron-%
gas ground-state energy locally. There are no empirical or fitted 
parameters, just references to general theoretical criteria.

Like composite molecules, adsorption systems have electrons in separate 
molecule-like regions, with exponentially decaying tails in between. Then 
the slowly varying electron gas, used in the original vdW-DF 
method~\cite{dion2004,thonhauser2007,rydberg2000,langreth2005}, 
might not be the most appropriate reference system for 
the gradient correction~\cite{LV1990}. 
Although promising results have been obtained for a variety of systems,
including adsorption \cite{langreth2009,Mats}, 
there is room for improvements.
For the recent vdW-DF2 functional, 
the gradient coefficient of the B88 exchange functional~\cite{B88} is 
used for the determination of the internal functional 
[Eq.~(12) of Ref.~\onlinecite{dion2004}] within the nonlocal 
correlation functional. This is based on application of the large-$N$ 
asymptote~\cite{Schwinger1980,Schwinger1981} on appropriate molecular 
systems. Using this method, Elliott and Burke~\cite{Elliott2009} 
have shown from first principles that the correct exchange gradient coefficient 
$\beta$ for an isolated atom (monomer) is essentially identical to the 
B88 value, which had been previously determined empirically~\cite{B88}.
Thus in the internal functional,
vdW-DF2~\cite{LeeEtAl10} replaces $Z_{ab}$ in that equation with
the value implied by the $\beta$ of B88. This procedure defines the
relationship between the kernels of vdW-DF and vdW-DF2 for the
nonlocal correlation energy.
Like vdW-DF, vdW-DF2 is a transferable functional
based on physical principles and approximation and without empirical input.

The vdW-DF method also needs to choose an overall exchange functional
to obtain the exchange contribution to the \textit{interaction} energy 
\textit{between} two monomers (for example). The original vdW-DF uses 
the revPBE~\cite{revPBE} exchange functional, good at separations in 
typical vdW complexes~\cite{dion2004,langreth2005,thonhauser2007}. 
The latter choice can be improved 
on~\cite{puzder2006,kannemann-becke2009,murray2009,klimes2010,cooper2010}. 
Recent studies suggest that the PW86 exchange functional~\cite{PW86} 
most closely duplicates Hartree-Fock interaction energies both for 
atoms~\cite{kannemann-becke2009} and molecules~\cite{murray2009}. 
The vdW-DF2 functional~\cite{LeeEtAl10} employs the 
PW86R functional~\cite{murray2009}, which more closely reproduces the 
PW86 integral form at lower densities than those considered by the 
original PW86 authors.

Evaluation of XC functionals with respect to other theoretical results 
is often done systematically, e.g., by benchmarking against the S22 
data sets~\cite{Jurecka2006,Sherrill2009,Molnar2009,Takatani2010,Szalewicz2010}.
The S22 sets have twenty-two prototypical 
small molecular duplexes for noncovalent interactions (hydrogen-bonded, 
dispersion-dominated, and mixed) in biological molecules and provides 
PECs at a very accurate level of 
wave-function methods, in particular the CCSD(T) method. 
However, by necessity, the electron systems in such sets 
have finite size. The original vdW-DF performs well on the S22 
dataset~\cite{Jurecka2006,Sherrill2009,Molnar2009,Takatani2010,Szalewicz2010}, 
except for hydrogen-bonded duplexes (underbinding 
by about 15\% \cite{langreth2009,LeeEtAl10}). The vdW-DF2 functional 
reduces the mean absolute deviations of binding energy and equilibrium 
separation significantly~\cite{LeeEtAl10}. Shapes of PECs away from 
the equilibrium separation are greatly improved. The long-range part 
of the vdW interaction, particularly crucial for extended systems, has 
a weaker attraction in the vdW-DF2, thus reducing the error to 8 meV at 
separations 1~{\AA} away from equilibrium~\cite{LeeEtAl10}. 

Experimental
information provides the ultimate basis for assessing
functionals. The vdW-DF functional has been promising in applications
to a variety of systems \cite{langreth2009}, but primarily
vdW bonded ones, typically tested on binding-energy
and/or bond-length values that happen to be available.
The vdW-DF2 functional has also been successfully applied to
some extended systems, like graphene and graphite \cite{LeeEtAl10},
metal-organic-frameworks systems \cite{LeeEtAl10B}, molecular crystal
systems \cite{MolcrysDF2}, physisorption
systems \cite{LeeEtAl10C,selfassembly}, 
liquid water \cite{mogelhoj} and layered oxides \cite{londero}. 
However, those studies are of
the common kind that focus on comparison against just
a few accessible observations. 

The key step taken by the present work is to benchmark
a full PEC in an extended system. Fortunately, for almost two
decades, accurate experimental values for the eigenenergies
of H$_2$ and D$_2$ molecules bound to Cu surfaces
\cite{andersson1993,perandersson1993} have 
been waiting for theoretical account and assessment.
The rich data bank covers results for
the whole shape of the physisorption potentials.

The H$_2$-Cu system is particularly demanding for the vdW-DF 
and vdW-DF2 functionals and alike. On one hand, H$_2$ is a 
small molecule with a large HOMO-LUMO gap, far from 
the low-frequency polarization modes assumed in derivation 
of these functionals \cite{dion2004,thonhauser2007,langreth2005,LeeEtAl10}. 
On the other hand, Cu is a metal, which 
has created particular concerns \cite{dobson}.

Chemically inert atoms and molecules adsorb physically on cold metal 
surfaces~\cite{persson2008}, with characteristic low desorption 
temperatures ranging from a few K (He) to tens of K (say Ar and CH$_4$). 
Adsorption energies, which take values from a few meV to around 100 meV, 
may be determined from thermal-desorption or isosteric-heat-of-adsorption 
measurements. For light adsorbates, like He and H$_2$, 
gas-surface-scattering experiments, involving resonance structure of 
the elastic backscattering, provide a more direct and elegant method, 
with accurate and detailed measurements of bound-level sequences in 
the potential well. The availability of isotopes with widely different 
masses (${}^3$He, ${}^4$He, H$_2$, D$_2$) permits a unique assignment 
of the levels and a determination of the well depth and ultimately 
a qualified test of model potentials~\cite{Roy}. 

The bound-level sequences of specific concern here were obtained using 
nozzle beams of para-H$_2$ and normal-D$_2$, i.e., the beams are 
predominantly composed of $j = 0$ molecules. This implies that the 
measured bound-state energies, $\epsilon_n$ (listed in Table~\ref{tab:1} 
for H$_2$ and D$_2$ on Cu(111)), refer to an isotropic distribution 
of the molecular orientation. For this particular 
ordering, all $\epsilon_n$ values fall accurately on a common curve when 
plotted versus the mass-reduced level number $\eta = (n + 1/2)/\sqrt{m}$. 
This implies a level assignment that is compatible with a single gas-surface 
potential for the two hydrogen isotopes~\cite{perandersson1993}. 
A third-order polynomial fit to the data yields for $\eta = 0$ a 
potential-well depth $D = 29.5$ meV. 

\begin{figure}[bth]
\centering
\includegraphics[width=8.5cm]{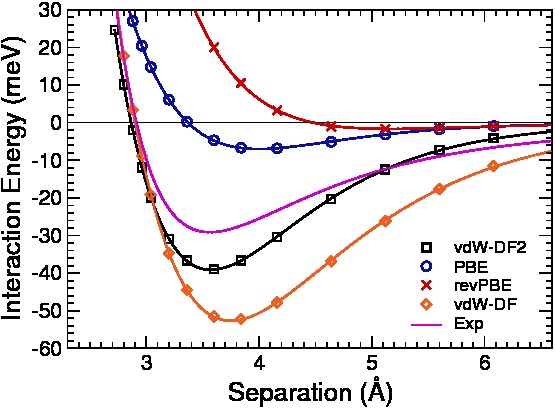}
\caption{Experimentally determined effective physisorption potential 
for H$_2$ on Cu(111)~\protect\cite{andersson1993},
compared with potential-energy curves for H$_2$ on Cu(111), calculated 
for the atop site in GGA-revPBE, GGA-PBE, 
vdW-DF2, and vdW-DF.}
\label{fig:1}
\end{figure}

The experimental energy levels in the H$_2$-Cu(111) PEC (see Table~\ref{tab:1})
may be
analyzed \cite{andersson1993,perandersson1993,persson2008} within the 
traditional theoretical picture 
\cite{zaremba1977,HarrisNordlander} of the 
interaction between inert adsorbates and metal surfaces: 
The PEC is then approximated as a 
superposition of a long-range vdW attraction, $V_{\mathrm{vdW}}$, 
and a short-range Pauli repulsion, $V_R$, reflecting overlap between 
tails of metal Bloch electrons and the adparticle's closed-shell 
electrons~\cite{zaremba1977,persson2008}.
This results in a laterally and angularly averaged potential 
$V_o(z) = V_R(z) + V_\mathrm{vdW}(z)$, where $z$ is the normal distance 
of the H$_2$ bond center from the jellium edge. 
The bound-level sequences in Table~\ref{tab:1} can be accurately 
reproduced ($< 0.3$ meV) by such a physisorption 
potential~\cite{perandersson1993,persson2008} (Fig.~\ref{fig:1}), 
having a well depth of 28.9 meV and a potential minimum
located 3.50~{\AA} outside the topmost layer of copper ion cores.
{}From the measured intensities of the first-order
diffraction beams, a very small lateral variation of the
H$_2$-Cu(111) potential can be deduced, $\sim 0.5$ meV at the
potential-well minimum.
       
A direct solution of the Schr{\"o}dinger equation 
in $V_o(z)$ reproduces the four low-energy eigenvalues      
to within 3\% of the measured ones.
It is therefore consistent \cite{saturation} 
to benchmark against the very accurately constructed 
experimental physisorption curve in Fig.~\ref{fig:1}.

\begin{table}
\begin{center}
\caption{\label{tab:1}Sequences of bound-state energies for H$_2$ 
and D$_2$ on Cu(111). DFT eigenvalues are calculated with the 
vdW-DF2 potential of Fig.~\ref{fig:1}, and experimental numbers 
are from \cite{andersson1993,perandersson1993}.}

\begin{tabular} {crrrr}
\ \ \ $n$ \ \ \ & \multicolumn{4}{c}{$\epsilon_n$ [meV]} \\
& \multicolumn{2}{c}{H$_2$} & \multicolumn{2}{c}{D$_2$}\\
 & \multicolumn{1}{c}{DFT}& 
\multicolumn{1}{c}{Exper.} & \multicolumn{1}{c}{DFT}
& \multicolumn{1}{c}{Exper.} \\
\hline
0  & $-32.6$ & $-23.9$  & $-34.4$  &            \\
1  & $-21.3$ & $-15.5$  & $-26.0$  & $-19.0$    \\
2  & $-12.1$ & $ -8.7$  & $-18.7$  & $-12.9$    \\
3  & $ -5.4$ & $ -5.0$  & $-12.4$  & $-8.9$     \\
4  &         &          & $ -7.4$  & $-5.6$     \\
5  &         &          & $ -3.5$  & $-3.3$    \\
\hline
\end{tabular}
\end{center}
\end{table}

Figure~\ref{fig:1} shows our comparison of density functional PECs against the
experimental physisorption potential for H$_2$ in an orientation above an
atop site on the Cu(111) surface. 
The PECs of Fig.~\ref{fig:1} are calculated with the 
vdW-DF~\cite{dion2004} and vdW-DF2 functionals~\cite{LeeEtAl10} as well as with
two generalized gradient approximations (GGA-PBE and GGA-revPBE). We
use an efficient vdW algorithm~\cite{FFTvdWDF} adapted from
SIESTA's \cite{siesta} vdW code within a modified version of the
plane-wave code ABINIT \cite{abinit}. The vdW interaction is treated
fully self-consistently \cite{thonhauser2007} (allowing also vdW forces
to relax the adsorption geometry). The computational costs are the same
with vdW-DF and vdW-DF2. Our choice of Troullier-Martins-type
norm-conserving pseudopotentials and a high cutoff energy (70 Ry)
ensures excellent convergence; the GPAW code~\cite{mortensen2005},
in its default mode, gives similar but less accurate 
results~\cite{kelkkanenThesis}.
We stress
that there is neither a damping nor saturation function
in vdW-DF and vdW-DF2 calculations. 
The need for an account of nonlocal correlations for the
description of vdW forces is illustrated by the GGA curves giving 
inadequate PECs.
The calculated well depth in vdW-DF, 53 meV, should 
be compared with the measured one, 29.5 meV \cite{perandersson1993}, and 
the one calculated from $V_o(z)$, suitably parametrized, 28.9 meV%
~\cite{andersson1993} (Fig.~\ref{fig:1}). We find that
the vdW-DF2 PEC lies close to the experimental physisorption potential, both 
at the equilibrium position and at separations further away from the 
surface. 

\begin{figure}[bt]
\centering
\includegraphics[width=8.5cm]{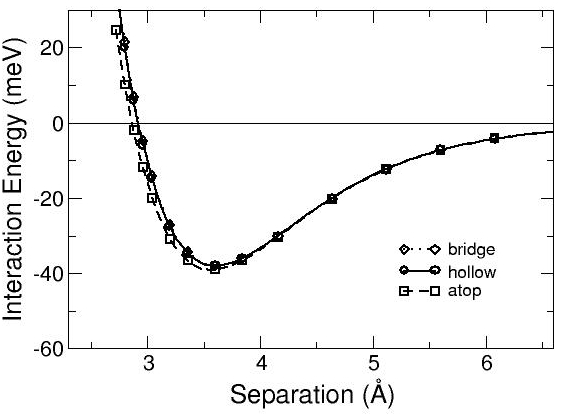}
\caption{Interaction potential for H$_2$ on Cu(111), calculated
self-consistently with the vdW-DF2 functional~\cite{LeeEtAl10}
in the bridge, hollow and atop sites.
}
\label{fig:2}
\end{figure}

Calculated PECs of 
H$_2$ above bridge, atop, and center sites on the Cu(111) surface are 
shown in Fig.~\ref{fig:2}, their closeness illustrating the lack of 
corrugation on this surface, as in experimental findings%
~\cite{andersson1993,perandersson1993}. Similar results for bridge, 
atop, and center sites on Cu(100) show the PECs on the two surfaces 
to be very close to each other, just like for the experimental result%
~\cite{andersson1993,perandersson1993}. The vdW-DF2 equilibrium 
separation is about 3.5 {\AA}, like the value that 
is deduced from 
the experimental data and for reasonable physical assumptions about the 
parameters~\cite{andersson1993,perandersson1993}. 

A further refined comparison is provided by the bound-state eigenvalues 
for the point-of-gravity motion of H$_2$ on Cu(111), both experimental 
data and those from the vdW-DF2 potential (Table~\ref{tab:1}).
In addition to PEC shapes and eigenenergy values, 
there should be comparison with values for well depth and equilibrium 
separation, experimental ones being (29.5 meV; 3.5 \AA) and those of the 
vdW-DF2 potential in Fig.~\ref{fig:1} 
(37 meV; 3.5 \AA). 
Viewing the facts that (i) vdW-DF2 is 
a first-principles method, where characteristic electron 
energies are typically in the eV range, and (ii) the test system and results 
are very demanding, as other popular methods deviate significantly more from 
the experimental curve (for instance, application of the DFT-D3(PBE)
method~\cite{grimme10}, with atom-pairwise specific dispersion coefficients 
and cutoff radii computed from first principles,  gives ($-88$ meV; 2.8 \AA) 
for the PEC minimum point).  We judge this as very promising.  So is the 
relative closeness of experimental and calculated eigenenergy values in 
Table~\ref{tab:1}. 

The discrepancies between the eigenvalues signal that the 
vdW-DF2 PEC might not have the right shape for H$_2$ on Cu(111). 
Reference~\onlinecite{LeeEtAl10} shows that  
vdW-DF2 benchmarks very well against the S22 data sets. 
It is possible that the metallic nature of the H$_2$/Cu(111)
system causes modifications in the details of the 
electrodynamical response; H$_2$/Cu(111) physisorption
constitutes a very strong challenge for the density functional.
For a well established conclusion, a more accurate theory 
is called for.

In summary, accurate and extensive experimental data
for the physisorption PEC of H$_2$ on Cu(111) are used
to evaluate vdW-DF functionals and their adequacy for
metal surfaces. The Cu(111) surface is chosen
here, as its flatness gives clarity in the analysis and
eliminates several side-issues that could have made interpretations
fuzzier. More generally, there exists an accurate data bank of
experimental physisorption information that challenges every density
functional approximation to produce relevant PECs. We propose that such
surface-related PEC benchmarking should find a broader usage,
supplementing, for example, S22 comparisons as an accelerator
in density-functional development. 

Several qualitative
similarities are found for both vdW-DF and vdW-DF2
functionals. The vdW-DF2 functional gives PECs
in a useful qualitative and quantitative agreement with
the experimental PEC, i.e.\ with respect to well depth,
equilibrium separation, and curvature of PEC near the
well bottom, and thus zero-point vibration frequency.
This is very promising for applications of this nonlocal
correlation functional at short and intermediate separations,
relevant for the adsorption. However, the accuracy
of experimental data is high enough to stimulate
a more detailed analysis of all aspects of the theoretical
description. This should be valuable for the further
XC-functional development.

The Swedish National Infrastructure for Computing (SNIC) is acknowledged for 
providing computer allocation and the Swedish Research Council for providing 
support to ES and PH. AK and BIL thank the Lundbeck foundation for sponsoring 
the center for Atomic-scale Materials Design and the Danish Center for 
Scientific Computing for providing computational resources. Work by KL and 
DCL is supported by NSF DMR-0801343.
Professor David Langreth was active in all aspects of the
research until his untimely death in May 2011. We would like to express
our sense both of personal loss and of loss
to our discipline.


\begin{thebibliography}{99}

\bibitem{Kohn} P. Hohenberg and W. Kohn, Phys. Rev. \textbf{136}, B864 (1964).

\bibitem{KohnSham} W. Kohn and L.J. Sham, Phys. Rev. \textbf{140}, A1133 (1965).

\bibitem{andersson1993}
S. Andersson and M. Persson, Phys. Rev. Lett. \textbf{70}, 202 (1993).

\bibitem{perandersson1993}
S. Andersson and M. Persson, Phys. Rev. B  \textbf{48}, 5685 (1993).

\bibitem{persson2008}
See, {\it e.g.}, M. Persson and S. Andersson, Chapter 4, ``Physi\-sorption Dynamics at Metal Surfaces", in
Handbook of Surface Science, Vol. 3 (Eds. E. Hasselbrink and B.I. Lundqvist), Elsevier,
Amsterdam (2008), p. 95.

\bibitem{Roy} R. J. Le Roy, Surf. Sci. \textbf{59}, 541 (1976).

\bibitem{dion2004} M. Dion, H. Rydberg, E. Schr\"oder, 
D.C. Langreth, and B.I. Lundqvist, Phys. Rev. Lett. 
\textbf{92}, 246401 (2004) and \textbf{95}, 109902(E) (2005). 

\bibitem{langreth2005} D.C. Langreth, M. Dion, H. Rydberg, E. Schr\"oder, 
P. Hyldgaard, and B.I. Lundqvist, Int. J. Quant. Chem. \textbf{101}, 599 (2005).

\bibitem{thonhauser2007} T. Thonhauser, V.R. Cooper, 
S. Li, A. Puzder, P. Hyldgaard, and D.C. Langreth,
Phys. Rev. B \textbf{76}, 125112 (2007).

\bibitem{LeeEtAl10} K. Lee, \'E.D. Murray, L. Kong, B.I. Lundqvist, and 
D.C. Langreth, Phys. Rev. B (RC) \textbf{82}, 081101 (2010).

\bibitem{PerdLangrI} D.C. Langreth and J.P. Perdew, Solid 
State Commun. \textbf{17}, 1425 (1975).

\bibitem{GunnLund} O. Gunnarsson and B.I. Lundqvist, Phys. Rev. B 
\textbf{13}, 4274 (1976).

\bibitem{PerdLangrII} D.C. Langreth and J.P. Perdew, Phys. Rev. B 
\textbf{15}, 2884 (1977).

\bibitem{rydberg2000} H. Rydberg, B.I. Lundqvist, D.C. Langreth, and M. Dion, 
Phys. Rev. B \textbf{62}, 6997 (2000).

\bibitem{LV1990} D.C. Langreth and S.H. Vosko, in 
``Density Functional Theory of Many-Fermion Systems", 
ed. S.B. Trickey, Academic Press, Orlando, 1990. 

\bibitem{langreth2009} D.C. Langreth, B.I. Lundqvist, 
S.D. Chakarova-K\"ack, V.R. Cooper, M. Dion, P. Hyldgaard, 
A. Kelkkanen, J. Kleis, L. Kong, S. Li, P.G. Moses, 
E. Murray, A. Puzder, H. Rydberg, E. Schr\"oder, and T. 
Thonhauser, J. Phys.: Cond. Mat. \textbf{21}, 084203 (2009).

\bibitem{Mats} Y.N. Zhang, F. Hanke, V. Bortolani, M. Persson, and R. Q. Wu,
Phys. Rev. Lett. {\bf 106}, 236103 (2011). 

\bibitem{B88} A.D. Becke, Phys. Rev. A \textbf{38}, 3098 (1988).

\bibitem{Schwinger1980} J. Schwinger, Phys. Rev. A 
\textbf{22}, 1827 (1980).

\bibitem{Schwinger1981} J. Schwinger, Phys. Rev. A 
\textbf{24}, 2353 (1981).

\bibitem{Elliott2009} P. Elliott and K. Burke, Can. J. Chem. 
\textbf{87}, 1485 (2009).

\bibitem{revPBE}
Y. Zhang and W. Yang, Phys. Rev. Lett. \textbf{80}, 890 (1998).

\bibitem{puzder2006} A. Puzder, M. Dion, and  D.C. Langreth, 
J. Chem. Phys. \textbf{126}, 164105 (2006).

\bibitem{kannemann-becke2009} F. O. Kannemann and A. D. Becke, J. Chem. Theory Comput.
\textbf{5}, 719 (2009).
 
\bibitem{murray2009} {\'E}.D. Murray, K. Lee, and D.C. Langreth,
Jour. Chem. Theor. Comput. \textbf{5}, 2754 (2009).

\bibitem{klimes2010} J. Klime\v{s}, D.R. Bowler, and A. Michaelides,
 J. Phys.: Condens. Matter \textbf{22}, 022201 (2010).

\bibitem{cooper2010} V.R. Cooper, Phys. Rev. B 
\textbf{81}, 161104(R) (2010).

\bibitem{PW86} J.P. Perdew and Y. Wang, Phys. Rev. B \textbf{33}, 8800(R) (1986).
 
\bibitem{Jurecka2006} P. Jure\v{c}ka, J. \v{S}poner, J. \v{C}ern\'y, and P. Hobza,
Phys. Chem. Chem. Phys. \textbf{8}, 1985 (2006).

\bibitem{Sherrill2009} D. Sherrill, T. Takatani, and E. G. Hohenstein, 
J. Phys. Chem. A \textbf{113}, 10146 (2009).

\bibitem{Molnar2009} L. F. Molnar, X. He, B. Wang, and K. M. Merz, J. Chem. 
Phys. \textbf{131}, 065102 (2009).

\bibitem{Takatani2010} 
T. Takatani, E.G. Hohenstein, M. Malagoli, M.S. Marshall, 
and C.D. Sherrill, J. Chem. Phys. \textbf{132}, 144104 (2010). 

\bibitem{Szalewicz2010} R. Podeszwa, K. Patkowski, and K. 
Szalewicz, Phys. Chem. Chem. Phys. \textbf{12}, 5974 (2010).

\bibitem{LeeEtAl10B} L.Z. Kong, G. Rom\'an-P\'erez, J.M. Soler, and D.C. Langreth, 
Phys. Rev. Lett. \textbf{103}, 096103 (2009).   

\bibitem{MolcrysDF2} K. Berland, {\O}. Borck, and P. Hyldgaard,
Comp. Phys. Commun. \textbf{182}, 1800 (2011). 

\bibitem{LeeEtAl10C} 
K. Lee, Y. Morikawa, and D.C. Langreth, Phys. Rev. B \textbf{82}, 155461 (2010).

\bibitem{selfassembly} J. Wyrick, D.-H. Kim, D. Sun, Z .Cheng, W. Lu, Y. Zhu, K. Berland, Y.S. Kim, E. Rotenberg, 
M. Luo, P. Hyldgaard, T.L. Einstein, and L. Bartels, Nano Letters \textbf{11}, 2944 (2011).

\bibitem{mogelhoj} A. M{\o}gelh{\o}j, A. Kelkkanen, K.T. Wikfeldt,
J. Schi{\o}tz, J.J. Mortensen, L.G.M. Pettersson, B.I. Lundqvist,
K.W. Jacobsen, A. Nilsson, and J.K. N{\o}rskov,
J. Phys. Chem. B (DOI: 10.1021/jp2040345), in print.

\bibitem{londero} E. Londero and E. Schr\"oder,
Computer Phys. Commun. \textbf{182}, 1805 (2011).

\bibitem{dobson} J.F. Dobson, Surf. Sci. \textbf{605}, 1621 (2011) and references therein.

\bibitem{zaremba1977}
E. Zaremba and W. Kohn, Phys. Rev. B \textbf{15}, 1769 (1977).

\bibitem{HarrisNordlander} P. Nordlander and J. Harris,
J. Phys. C \textbf{17}, 1141 (1984).

\bibitem{saturation} We note that $V_{\rm vdW}(z)$ in the potential 
$V_o(z)$ involves a prefactor $f(z)$, which introduces a
saturation of the attraction at atomic-scale separations.
The function $f(z)$ lacks a rigorous prescription and this 
results in a level of arbitrariness to $V_{\rm vdW}(z)$.
It should be noted, though, that the saturation function $f(z)$ of 
Refs.\ \protect\onlinecite{HarrisNordlander,andersson1993,perandersson1993,persson2008}
has a formal resemblance to the damping functions 
introduced to adjust for double counting in 
semi-empirical DFT descriptions \protect\cite{grimme10}
of dispersive interactions.  We emphasize that $f(z)$ of $V_{\rm vdW}(z)$  
is likely to work across several length scales \protect\cite{Berland2010}
because $V_{\rm vdW}(z)$ includes the effects of image planes and 
is consistent with the Zaremba-Kohn formulation of 
physisorption \protect\cite{zaremba1977}.
	
\bibitem{FFTvdWDF} G. Rom\'an-P\'erez and J.M. Soler, Phys. Rev. Lett. 
\textbf{103}, 096102 (2009).

\bibitem{siesta}
P. Ordej\'on, E. Artacho, and J.M. Soler, Phys. Rev. \textbf{53}, 10441(R) (1996);
J.M. Soler, E. Artacho, J.D. Gale, A. Garc\'ia, J. Junquera, P. Ordej\'on, and
D. S\'anchez-Portal, J. Phys.: Condens. Matter \textbf{14}, 2745 (2002).

\bibitem{abinit}
X. Gonze, J.-M. Beuken, R. Caracas, F. Detraux, M. Fuchs, G.-M. Rignanese,
L. Sindic, M. Verstraete, G. Zerah, F. Jollet, M. Torrent, A. Roy, M. Mikami,
Ph. Ghosez, J.-Y. Raty, and D.C. Allan, Comput. Mater. Sci. \textbf{25}, 478 (2002).

\bibitem{mortensen2005} J.J. Mortensen, L.B. Hansen, and K.W.
Jacobsen, Phys. Rev. B \textbf{71}, 035109 (2005).

\bibitem{kelkkanenThesis} A. Kelkkanen, 
\textit{Implementation of van der Waals Forces in DFT for 
Nanoelectronics and Water Structure}, Thesis (CAMd, DTU, 2011).

\bibitem{grimme10} S. Grimme, J. Antony, S. Ehrlich, and H. Krieg, J. Chem. Phys. \textbf{132}, 154104 (2010).

\bibitem{Berland2010} K. Berland and P. Hyldgaard, J. Chem. Phys. \textbf{132}, 134705 (2010).


\end{thebibliography}
\end{document}